\documentclass[twocolumn]{aastex631}

\begin{document}

\shortauthors{Luhman et al.}

\shorttitle{JWST/NIRSpec Observations of Coldest Known Brown Dwarf}

\title{JWST/NIRSpec Observations of the Coldest Known Brown 
Dwarf\footnote{Based on observations made with the NASA/ESA/CSA 
James Webb Space Telescope.}}

\author{K. L. Luhman}
\affiliation{Department of Astronomy and Astrophysics,
The Pennsylvania State University, University Park, PA 16802, USA;
kll207@psu.edu}
\affiliation{Center for Exoplanets and Habitable Worlds, 
The Pennsylvania State University, University Park, PA 16802, USA}

\author{P. Tremblin}
\affiliation{Universit\'{e} Paris-Saclay, UVSQ, CNRS, CEA, 
Maison de la Simulation, F-91191, Gif-sur-Yvette, France}

\author{C. Alves de Oliveira}
\affiliation{European Space Agency, European Space Astronomy Centre, Camino 
Bajo del Castillo s/n, E-28692 Villanueva de la Ca\~{n}ada, Madrid, Spain}

\author{S. M. Birkmann}
\affiliation{European Space Agency (ESA), ESA Office, Space Telescope
Science Institute, 3700 San Martin Drive, Baltimore, MD 21218, USA}

\author{I. Baraffe}
\affiliation{Physics \& Astronomy Dpt, University of Exeter, Exeter
EX4 4QL, UK}

\author{G. Chabrier}
\affiliation{Physics \& Astronomy Dpt, University of Exeter, Exeter EX4 4QL, 
UK}
\affiliation{Ecole normale sup\'{e}rieure de Lyon, CRAL, CNRS UMR 5574,
F-69364, Lyon Cedex 07, France}

\author{E. Manjavacas}
\affiliation{AURA for the European Space Agency, Space Telescope Science
Institute, 3700 San Martin Drive, Baltimore, MD 21218, USA}
\affiliation{Department of Physics \& Astronomy, Johns Hopkins University,
Baltimore, MD 21218, USA}

\author{R. J. Parker}
\affiliation{Department of Physics and Astronomy, The University of Sheffield,
Hicks Building, Hounsfield Road, Sheffield S3 7RH, UK}
\affiliation{Royal Society Dorothy Hodgkin Fellow}

\author{J. Valenti}
\affiliation{Space Telescope Science Institute, 3700 San Martin Drive, 
Baltimore, MD 21218, USA}

\begin{abstract}

We present 1--5~\micron\ spectroscopy of the coldest known brown dwarf,
WISE J085510.83$-$071442.5 (WISE 0855), performed with the Near-Infrared
Spectrograph (NIRSpec) on board the James Webb Space Telescope (JWST).
NIRSpec has dramatically improved the
measurement of spectral energy distribution of WISE 0855 in terms of
wavelength coverage, signal-to-noise ratios, and spectral resolution.
We have performed preliminary modeling of the NIRSpec data
using the {\tt ATMO 2020} models of cloudless atmospheres, arriving
at a best fitting model that has $T_{\rm eff}=285$~K.
That temperature is $\sim$20~K higher than the value derived by combining
our luminosity estimate with evolutionary models (i.e., the radius in the model
fit to the SED is somewhat smaller than expected from evolutionary models).
Through comparisons to the model spectra, we detect
absorption in the fundamental band of CO, which is consistent with an
earlier detection in a ground-based spectrum and indicates the presence
of vertical mixing. Although PH$_3$ is expected in Y dwarfs that experience 
vertical mixing, it is not detected in WISE 0855. 
Previous ground-based $M$-band spectroscopy of WISE 0855
has been cited for evidence of H$_2$O ice clouds, but we find that the NIRSpec
data in that wavelength range are matched well by our cloudless model.
Thus, clear evidence of H$_2$O ice clouds in WISE 0855 has not been
identified yet, but it may still be present in the NIRSpec data.
The physical properties of WISE 0855, including the presence of H$_2$O clouds,
can be better constrained by more detailed fitting with both cloudless
and cloudy models and the incorporation of unpublished 5--28~\micron\ data
from the Mid-infrared Instrument on JWST.

\end{abstract}

\section{Introduction}
\label{sec:intro}

Y is the coldest spectral class of brown dwarfs 
\citep[$T_{\rm eff}\lesssim$500 K,][]{cus11,kir21}.
Because the near-infrared (IR) fluxes of brown dwarfs collapse from late T into
the Y class \citep{kir11}, sensitive space-based telescopes operating at
mid-IR wavelengths offer the best opportunity for detecting
Y dwarfs. Given their extremely low luminosities, detections of Y dwarfs
are restricted to the solar neighborhood, whose members are scattered
in all directions in the sky. As a result, the all-sky mid-IR survey performed
by the Wide-field Infrared Survey Explorer \citep[WISE,][]{wri10} has
uncovered most of the $\sim50$ objects that are known or suspected to be
Y dwarfs \citep{cus11,cus14,kir12,kir13,tin12,tin18,luh14b,pin14,sch15,mar19,mar20,bar20,mei20a,mei20b,rob23}\footnote{The photometry of Ross~19B indicates that it is near the T/Y boundary \citep{sch21,mei23}.}, although a few have been found 
through companion surveys with the Spitzer Space Telescope 
\citep{wer04,luh11,luh12}, ground-based near-IR adaptive optics imaging 
\citep{liu11,liu12,dup15}, and the James Webb Space Telescope 
\citep[JWST,][]{gar23,cal23}.
These discoveries have been accompanied by the development of atmospheric
and evolutionary models for the coldest brown dwarfs and giant planets, which
can be tested with observations of Y dwarfs
\citep{bur03,sau08,mor12,mor14a,mor14b,mor18,sau12,tre15,zal19,phi20,tan21,mar21,man22,muk22,lac23}\footnote{\citet{leg21} and \citet{mei23}
included modified versions of the models from \citet{tre15} and \citet{phi20}.}.

The coldest known brown dwarf is WISE J085510.83$-$071442.5 (hereafter WISE 
0855). \citet{luh14a} and \citet{kir14} identified it as a high proper
motion object using two epochs of images from WISE.  By obtaining images at 
two additional epochs with Spitzer and measuring its parallax, \citet{luh14b}
demonstrated that WISE 0855 is a nearby brown dwarf (the fourth
closest known system) and estimated a temperature of $\sim$250~K from 
its absolute magnitude at 4.5~\micron. The distance of WISE 0855 has been 
further refined through continued astrometric monitoring with Spitzer, the 
reactivated WISE mission \citep[NEOWISE,][]{mai14}, and the Hubble Space 
Telescope \citep[2.278$\pm$0.012~pc,][]{luh14pi,luh16,wri14,kir19,kir21}.
In an effort to measure its spectral energy distribution (SED) for comparison
to models of Y dwarfs, deep optical and near-IR imaging has been performed
on WISE 0855 with Hubble and large ground-based telescopes
\citep{bea14,fah14,kop14,luh14b,wri14,luh16,sch16,zap16,leg17}.
Near-IR spectroscopy of WISE 0855 is not feasible with 
these facilities, but through very long exposure times with Gemini Observatory,
\citet{ske16} and \citet{mor18} were able to obtain spectra 
within portions of the mid-IR (the $L$ and $M$ bands) where the atmosphere 
is not opaque.

JWST is the most sensitive IR telescope deployed to date and operates across 
a wavelength range that should encompass most of the flux emitted by Y dwarfs. 
Therefore, it is the ideal facility for measuring the SEDs of Y dwarfs
like WISE 0855 \citep{bei23}.
In fact, given its status as both the coldest known brown dwarf and the
fourth closest known system and the limited ability of other telescopes
to observe it, WISE 0855 is one of the most appealing targets of any kind for 
JWST.  In this paper, we present spectroscopy of WISE 0855 from 
0.8--5.5~\micron\ performed with the Near-Infrared Spectrograph on 
JWST \citep[NIRSpec,][]{jak22}.

\section{Observations}

WISE 0855 was observed with NIRSpec through guaranteed time observation 
(GTO) program 1230 (PI: C. Alves de Oliveira) on 2023 April 17 (UT).
After slewing to WISE 0855, wide aperture target acquisition (WATA)
was performed, which employs the S1600A1 aperture ($1\farcs6\times1\farcs6$).
We used a combination of filter (CLEAR), readout pattern (NRSRAPID), and 
subarray (SUB2048) that would produce a suitable signal-to-noise ratio (SNR)
for acquisition given the SED of WISE 0855 \citep{luh16}. 
Through WATA, the centroid of the target was measured, a small angle 
maneuver was performed to center it in S1600A1, and an offset
was applied to place the target in the S200A1 fixed slit, which
has an angular size of $3\farcs2\times0\farcs2$.
One set of data was collected with the G395M grating, the F290LP filter,
the NRSRAPID readout pattern, 840 groups per integration, one integration
per nod, and three nod positions.
A second data set was taken with the PRISM disperser, the CLEAR filter,
the NRSRAPID readout pattern, 1083 groups per integration, three integrations 
per nod, and three nod positions. Both observations employed the SUBS200A1
subarray. The total exposure times with these configurations were 3931 
and 15,200~s, respectively.

\section{Data Reduction}
\label{sec:reduction}

The data reduction began with the retrieval of the {\tt uncal} files from the
Mikulksi Archive for Space Telescopes (MAST):
\dataset[doi:10.17909/1fsf-6j64]{http://dx.doi.org/10.17909/1fsf-6j64}.
We performed ramps-to-slopes processing on those files using 
a custom pipeline developed by the ESA NIRSpec science operations team, 
which adopts the same algorithms included in the official STScI pipeline
\citep{alv18,fer22}, except for a correction for "snowballs" \citep{bok23}
and a correction for residual correlated noise that is based on unilluminated 
pixels in the subarray outside the spectral trace \citep{esp23}.
For the PRISM data, which had multiple integrations, we adopted the median 
of the individual integration count rate maps at each nod position. 
The median count rate map for each nod was background subtracted using 
the averaged rate maps of the other two nods and flat fielded.
Spectra were extracted for the individual nod positions and combined 
with outlier rejection.  The same processing steps were performed for 
standard stars Gaia DR3 2260019315938461952 and Gaia
DR3 1634280312200704768 in G395M and PRISM, respectively, which were observed
during JWST commissioning through program 1128 (PI: N. L\"{u}tzgendorf).
The spectra of these stars and their flux models
\citep{boh14}\footnote{\url{https://www.stsci.edu/hst/instrumentation/reference-data-for-calibration-and-tools/astronomical-catalogs/calspec}}
were used for an initial flux calibration of the spectra of WISE 0855. 
The flux calibrations were then adjusted so that they agree with photometry
from Spitzer (Section~\ref{sec:comp}).
The reduced spectra have sufficient SNRs to be useful at 
2.87--5.09 and 5.37--5.54~\micron\ (G395M) and 0.77--5.53~\micron\ (PRISM).
The gap in the G395M data is from the separation between the two detector 
arrays. Our reduction yielded the widest possible wavelength coverage, 
extending slightly beyond the nominal ``science range" officially supported
by the STScI pipeline, while ensuring spectrophotometric calibration.
The spectral resolution is $\sim$1000 for G395M and ranges from 
$\sim40$ near 1.1~\micron\ to $\sim$300 at the longest wavelengths
for the PRISM disperser.

\section{Analysis}

\subsection{Comparison to Previous Observations}
\label{sec:comp}

We compare the NIRSpec data to previous observations of WISE 0855
between 1--5~\micron. We first consider the photometric measurements with the
smallest errors, consisting of data in the following filters: 
the F105W, F110W, F127M, and F160W bands on Hubble's Wide Field Camera 3 
\citep{kim08} measured by \citet{luh16} and \citet{sch16},
the [3.6] and [4.5] bands of Spitzer measured by \citet{luh16}, and W2 
from the NEOWISE-R Single Exposure Source Table \citep{cut23},
which contains 18 epochs for WISE 0855 between 2014 and 2022.
Each of the NEOWISE epochs consisted of $\sim10$--20 measurements that span
a few days. For each epoch, we calculated the median and standard error of W2.
We exclude the two epochs from the original WISE survey in 2010 since the 
photometry of WISE 0855 was contaminated by emission from
background stars \citep{luh14b}.
For bands that have multiple epochs of photometry with the exception of [3.6], 
we adopt the midpoint of the measurements and an uncertainty that is 
represented by the range spanned by those measurements and their errors.
The mean, median, and midpoint are nearly identical for [4.5], and the same
is true for W2 as well. The adopted values of [4.5] and W2 are 13.87$\pm$0.08 
and 13.90$\pm$0.16, respectively.
Since [3.6] was measured in only four of the nine [4.5] epochs from 
\citet{luh16}, we have calculated the weighted mean of the $[3.6]-[4.5]$ colors
among those four epochs (3.49) and applied it to the adopted measurement of 
[4.5] to arrive at our adopted value for [3.6]. The uncertainty for [4.5] 
is also assigned to [3.6]. 

We have calculated synthetic Vega magnitudes in the aforementioned bands
from the NIRSpec PRISM spectrum of WISE 0855.
In addition, photometry has been derived from the G395M spectrum in the
two bands that it fully covers, [3.6] and [4.5]. 
These calculations were performed by convolving the spectra of WISE 0855 and 
a spectrum of Sirius \citep{rie23} with the photon-counting spectral responses 
of the filters \citep{hor08,jar11,kri21}\footnote{\url{https://www.stsci.edu/hst/instrumentation/wfc3/performance/throughputs}}, integrating the resulting fluxes in photons, and 
assuming that Sirius has a magnitude of $-$1.395 in each band.

\begin{figure}
\epsscale{1.1}
\plotone{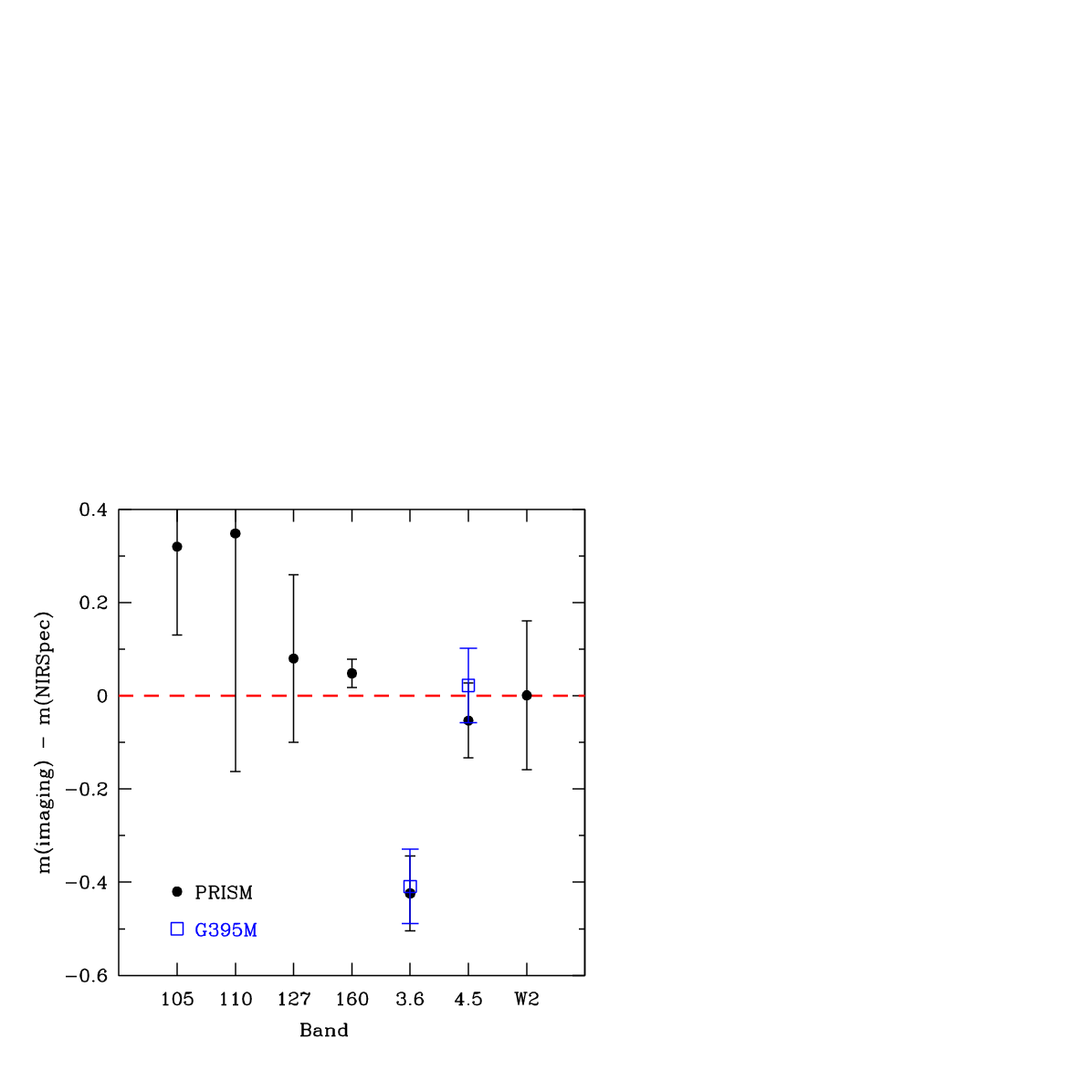}
\caption{Differences between previous photometry from images of
WISE 0855 and synthetic photometry measured from NIRSpec PRISM
and G395M spectra.}
\label{fig:dm}
\end{figure}

In Figure~\ref{fig:dm}, we have plotted the differences between 
the imaging and synthetic photometry. For most bands, the synthetic
photometry falls within the range of previous measurements.  The primary 
exception is [3.6], which is much fainter ($\sim$0.4~mag) in both NIRSpec
spectra than in the Spitzer images. The spectra and imaging agree more closely 
in [4.5], which means that $[3.6]-[4.5]$ is redder in the spectra than in 
the images. To further investigate this discrepancy, we have derived
synthetic [3.6] and [4.5] for the one other Y dwarf with an available
NIRSpec PRISM spectrum, WISE J035934.06$-$540154.6 
\citep[hereafter WISE 0359,][]{bei23}. 
We find that its color from NIRSpec is 0.28~mag redder than the value 
measured with Spitzer \citep{kir12}, resembling the result for WISE 0855. 
We also have compared the synthetic and imaging colors 
in W2$-$[4.5] for these Y dwarfs. The PRISM data indicate 
W2$-[4.5]=-0.02$ for both WISE 0855 and WISE 0359
whereas Spitzer and WISE/NEOWISE have measured 0.03 and 0.06, respectively.
It would be useful to perform similar comparisons for a larger 
sample of T and Y dwarfs observed by NIRSpec.

The differences between synthetic and imaging values of $[3.6]-[4.5]$ 
and W2$-$[4.5] for WISE 0855 and WISE 0359 could be caused by errors in the 
filter response functions or variability, although neither Y dwarf has shown 
large enough variability to explain the discrepancies in [3.6], as discussed 
earlier in this section and elsewhere \citep{esp16,bro23}. 
Synthetic [3.6] photometry for T and Y dwarfs is particularly sensitive to 
errors in the response function since their fluxes vary significantly across 
the bandpass, which overlaps with the fundamental band of CH$_4$. 
The response functions for the IRAC filters do shift slightly in wavelength
with position on the detector arrays \citep{hor08}. For the [3.6] band, the 
maximum redward shift relative to the average response function that we have 
adopted is $\sim80$~\AA, but we find that a shift of $\sim500$~\AA\ would be 
needed to account for the discrepancy in the photometry.
An error in the response function could partially
explain why synthetic values of $[3.6]-[4.5]$ from model spectra
for Y dwarfs are redder than observed colors \citep{leg21}.

Among the bands that we have discussed, [4.5] is the best one for
refining the flux calibrations of the PRISM and G395M spectra of
WISE 0855 since it contains a large fraction of the flux in 
those data, is fully encompassed by both spectra, and has the most
accurate photometry available from previous imaging.
Based on their synthetic photometry, the PRISM spectrum is 0.053~mag fainter 
and the G395M spectrum is 0.022~mag brighter than our adopted photometry 
of [4.5]=13.87. We have elected to multiply the spectra by the appropriate
factors to bring them into agreement with the latter.
Doing so results in a modest discrepancy of $\sim$0.1~mag between synthetic
and imaging photometry in F160W, which could be explained by variability.

\begin{figure}
\epsscale{1.2}
\plotone{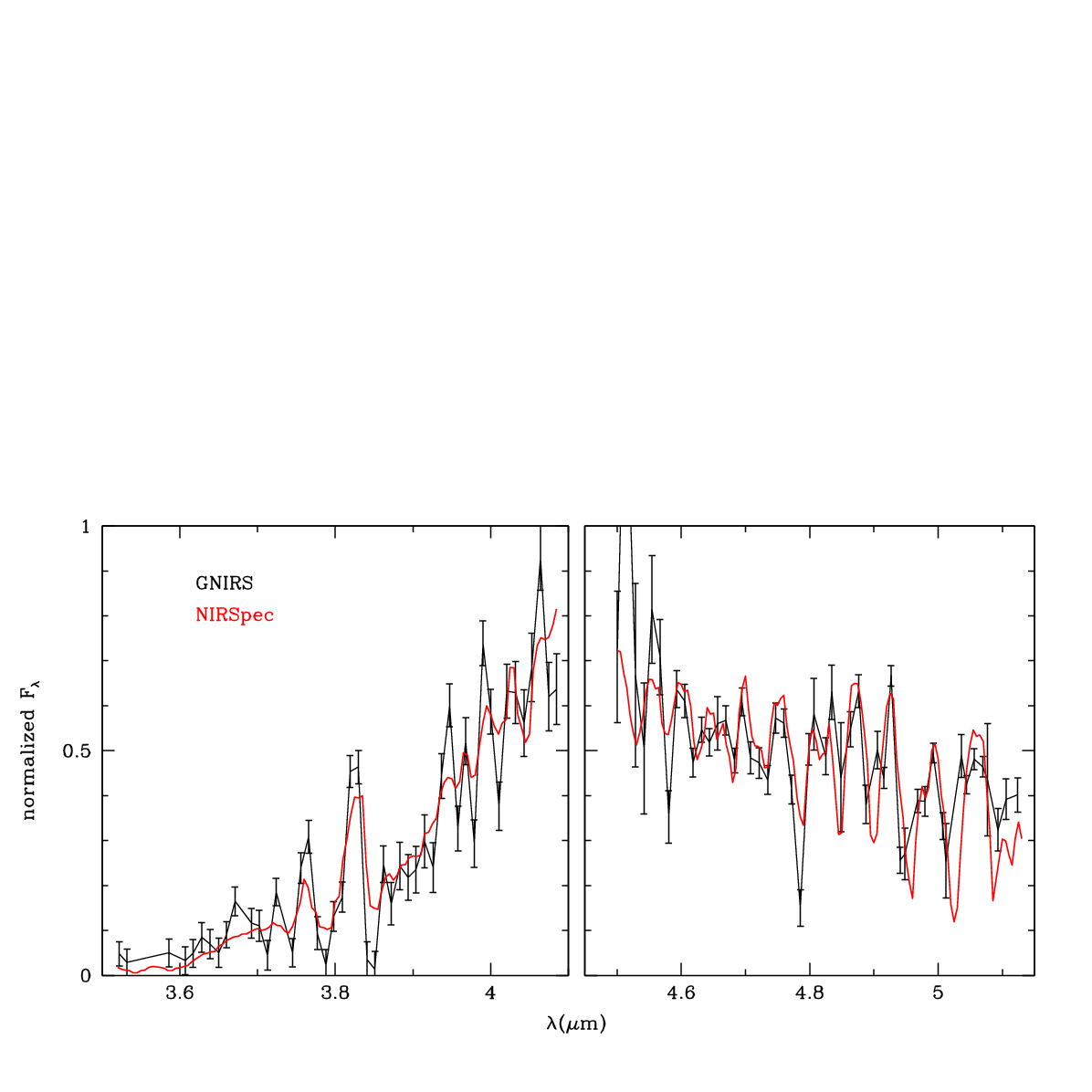}
\caption{Comparison of spectra for WISE 0855 from 
Gemini/GNIRS \citep{ske16,mor18,mil20} and JWST/NIRSpec (this work).}
\label{fig:gem}
\end{figure}

\citet{ske16} and \citet{mor18} obtained $M$ and $L$-band spectra, 
respectively, of WISE 0855 with the Gemini Near-Infrared Spectrograph 
\citep[GNIRS,][]{eli06}.  An updated reduction of the data from \citet{ske16}
was presented by \citet{mil20}. The data were binned to resolutions of 
$\sim250$ ($L$) and $\sim300$ ($M$), after which they had median SNRs of 
$\sim$5 ($L$) and 13 ($M$).
In Figure~\ref{fig:gem}, we compare the binned spectra from \citet{mor18} and 
\citet{mil20} to the NIRSpec PRISM spectrum, which has a similar resolution. 
The errors in the NIRSpec fluxes are not plotted since most are too small to 
be distinguishable from the spectrum. The GNIRS and NIRSpec spectra have 
similar spectral slopes and show many of the same absorption features, 
although the strengths of some of the features differ by substantially 
more than 1~$\sigma$.

\subsection{Spectral Features}
\label{sec:features}

\begin{figure}
\epsscale{0.9}
\plotone{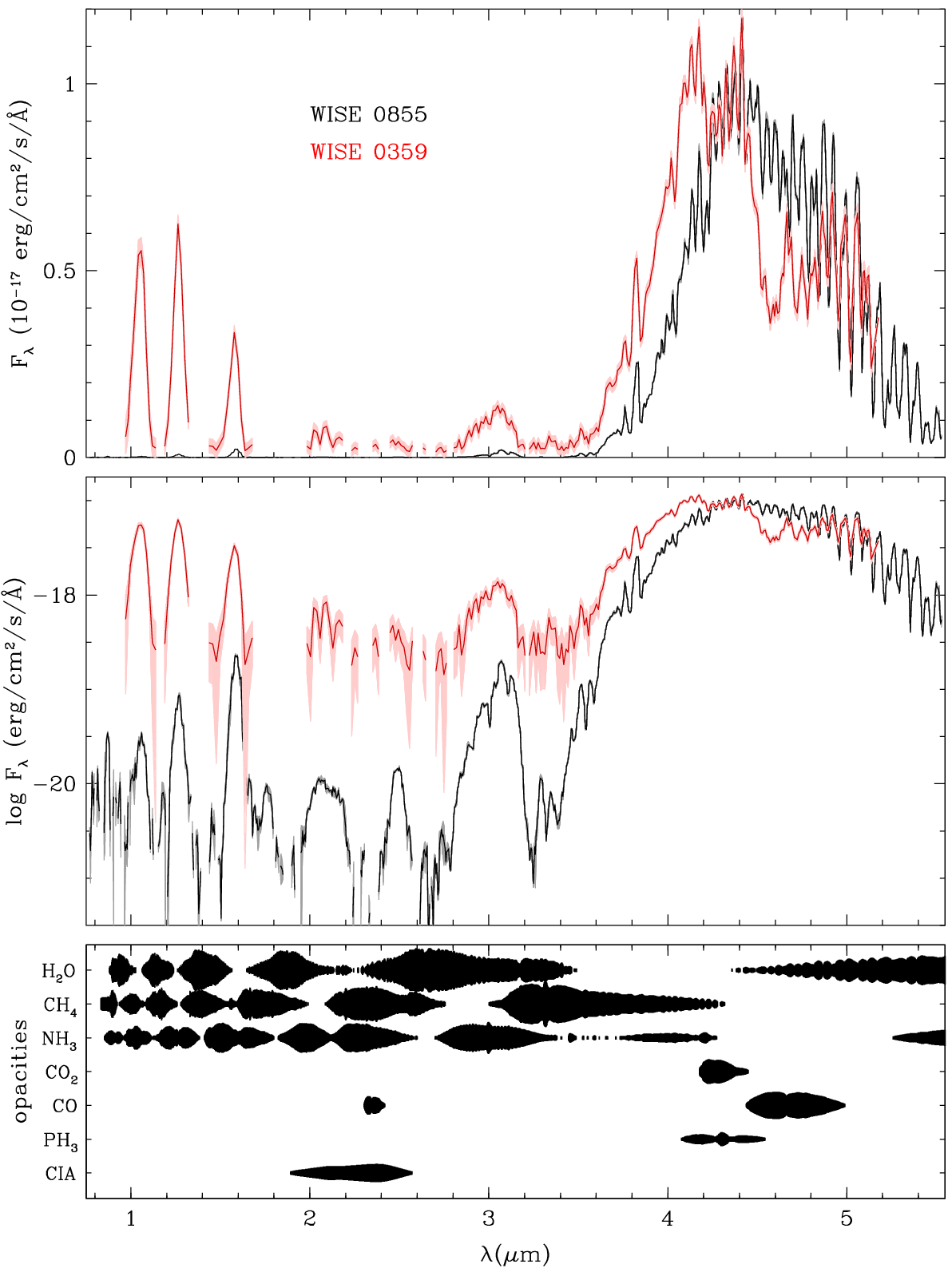}
\caption{NIRSpec PRISM spectra of WISE 0855 (this work) and 
the Y0 dwarf WISE 0359 \citep{bei23} on linear and logarithmic flux scales
(top and middle). The spectrum of WISE 0359 has been scaled to roughly
match the data for WISE 0855 at 4--5~\micron.
Uncertainties at $\pm1\sigma$ are plotted as gray and pink bands. 
Fluxes that are consistent with zero ($F_\lambda-\sigma\leq0$) have been 
omitted. The bottom panel shows a representation of opacities 
in which the vertical thickness of each band is proportional to 
the logarithm of the abundance-weighted absorption cross section for a given
molecule at P=1~bar and $T_{\rm eff}=250$~K (bottom). The opacities of CO$_2$, 
CO, and PH$_3$ have has been scaled by factors of 1000, 100, and 1000, 
respectively, so that their strongest features are visible.}
\label{fig:spec1}
\end{figure}

\begin{figure}
\epsscale{1}
\plotone{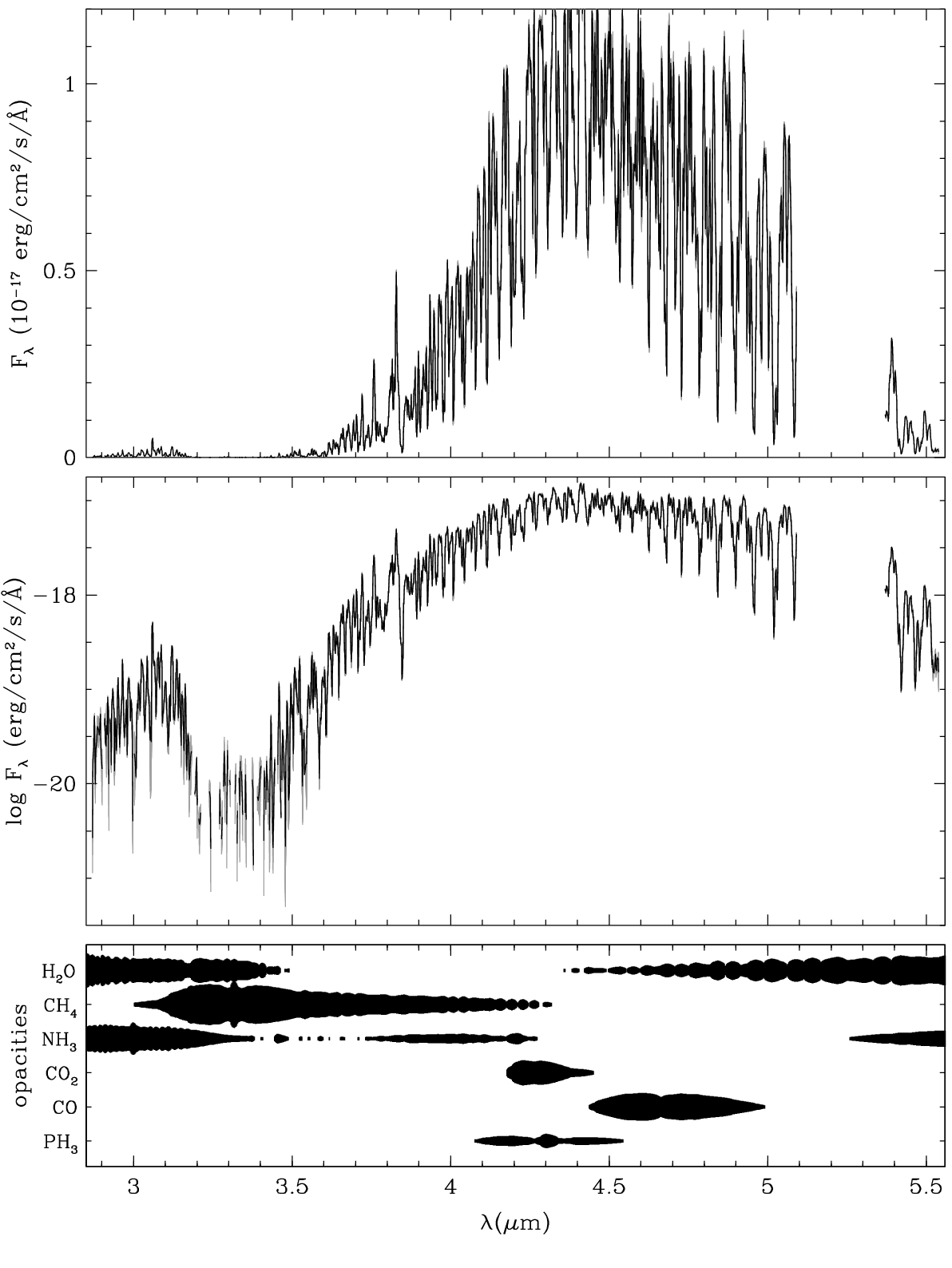}
\caption{NIRSpec G395M spectrum of WISE 0855 on linear and logarithmic flux 
scales (top and middle). Uncertainties at $\pm1\sigma$ are plotted as gray
bands. Fluxes that are consistent with zero ($F_\lambda-\sigma\leq0$) have been
omitted. The bottom panel shows the opacities from Figure~\ref{fig:spec1}.}
\label{fig:spec2}
\end{figure}

The NIRSpec spectra of WISE 0855 are presented in Figures~\ref{fig:spec1} 
and \ref{fig:spec2} (PRISM and G395M).  The spectra are available in
electronic files associated with those figures. Since the fluxes span a very 
large range, the spectra are shown on both linear and logarithmic flux scales. 
Fluxes that are consistent with zero at 1~$\sigma$ are not plotted.
Given that the fluxes vary significantly with wavelength, the SNRs
do so as well. In the fainter half of the PRISM spectrum ($<$2.5~\micron), 
the brightest 50\% of pixels have a median SNR of $\sim20$ while no significant
flux is detected in the strongest absorption bands. At the wavelengths
of that half of the PRISM spectrum,
only photometry (usually with low SNR) has been available from previous 
observations (Section~\ref{sec:comp}). The brightest fluxes in the PRISM
data are at 4--5~\micron, where the median SNR is $\sim90$.
Given their SNRs, wavelength coverage, and spectral resolution (for G395M),
the NIRSpec data have provided a dramatic improvement in the characterization
of the SED of WISE 0855.

The SED of WISE 0855 is shaped by absorption from several molecular species.
In the bottom panels of Figures~\ref{fig:spec1} and \ref{fig:spec2},
we have plotted a representation of the opacities for the dominant
absorbers that are expected in Y dwarfs, consisting of H$_2$O, CH$_4$, 
NH$_3$, CO, CO$_2$, and PH$_3$ and collision-induced absorption (CIA)
of H$_2$ and He.  In those opacity diagrams, 
the vertical thickness of each band is proportional to 
the logarithm of the abundance-weighted absorption cross section 
for a given molecule at P=1 bar and $T_{\rm eff}=250$~K based on the model of 
WISE 0855 from \citet{leg21}. We have summed the abundance-weighted cross 
sections for the CIA absorption of H$_2$ and He.
For the model in question, the opacities of CO$_2$, CO, and PH$_3$ are
much smaller than those of the other species, so for the purposes of
Figures~\ref{fig:spec1} and \ref{fig:spec2},
we have scaled the opacities of CO$_2$, CO, and PH$_3$ by factors of 1000, 
100, and 1000, respectively.

For comparison to WISE 0855, we have included in Figure~\ref{fig:spec1}
the NIRSpec PRISM spectrum of WISE 0359 \citep{bei23}, which has a 
spectral type of Y0 \citep{kir12}. The data for WISE 0359 have been
scaled to roughly match the fluxes of WISE 0855 at 4--5~\micron.
WISE 0359 is bluer from near- to mid-IR wavelengths than WISE 0855,
which reflects its earlier spectral type and higher temperature
\citep[$\sim$450~K,][]{bei23}. WISE 0359 has strong absorption 
from CO$_2$ and CO at 4--5~\micron, whereas WISE 0855 lacks CO$_2$ and
has much weaker CO \citep[][Section~\ref{sec:models}]{mil20}.
\citet{bei23} noted the detection of a relatively narrow feature
at 3~\micron\ in WISE 0359, which they attributed to the Q branch of the 
$\nu_1$ band of NH$_3$. That feature is detected in WISE 0855 as well.

\subsection{Spectral Classification}

Most known Y dwarfs have been classified as Y0 or Y1 \citep{kir21}.
One Y dwarf, WISEPA J182831.08+265037.8, has a spectral type of $\geq$Y2
\citep{kir12}. WISE 0855 is the one remaining Y dwarf, 
which has been assigned a tentative classification of Y4 based on
its colors and absolute magnitudes \citep{kir19,kir21}. NIRSpec has provided 
the first spectrum of WISE 0855 that is suitable for spectral
classification. Those data are consistent with an effective temperature
that is significantly cooler than that of other known Y dwarfs
(Figure~\ref{fig:spec1}, Section~\ref{sec:models}), so Y4 remains a reasonable
choice for its spectral type. The definition of classes later than Y1 will
require JWST spectra of a larger sample of the coolest known Y dwarfs 
(e.g., JWST program 2302, PI: M. Cushing) and would be facilitated by the 
discovery of objects that more closely approach the temperature of WISE 0855.

\subsection{Previous Comparisons to Models}

Previous studies have compared photometry and spectroscopy of WISE 0855
to the predictions of atmospheric models in an attempt to characterize 
its atmosphere and test the models. Such models are categorized in part
based on whether they assume chemical equilibrium or nonequilibrium 
chemistry due to vertical mixing, and whether they assume clear or cloudy
atmospheres. Warmer Y dwarfs may have sulfide clouds \citep{mor12} and
show evidence of nonequilibrium chemistry \citep{cus11,mil20,leg21,bei23}.
A brown dwarf as cold as WISE 0855 is expected to have clouds of H$_2$O
ice \citep[$<$350 K,][]{bur03,mor14a} and could experience nonequilibrium 
chemistry given its presence in the atmosphere of Jupiter \citep{pri75,vis06}.

Early studies of WISE 0855 found that its photometric SED was poorly matched
by all of the models from available grids and did not favor one model
prescription over another 
\citep[e.g., cloudless vs. cloudy,][]{luh14pi,luh16,sch16,leg17}.
Photometric monitoring with Spitzer has measured the mid-IR variability
of WISE 0855, which is similar to that of warmer Y dwarfs and is
inconclusive regarding the presence of H$_2$O ice clouds \citep{esp16}.
\citet{ske16}, \citet{mor18}, and \citet{mil20} found that cloudy
models provided a better match to their $M$-band spectra than
cloudless models, which was cited as evidence of H$_2$O clouds. 
However, given the untested nature of models in the temperature regime of
WISE 0855, if one model prescription appears to be favored by data in
a limited range of wavelengths, the same may not be true at other wavelengths
or with other suites of models. Indeed, the cloudy models from those studies 
were discrepant in multiple bands of the photometric SED 
while \citet{leg21} were able to use cloudless models
\citep{phi20} to simultaneously reproduce the slopes of the $L/M$-band
spectra and provide a good fit to the SED.
Meanwhile, detections of PH$_3$ and CO would be two of the most obvious
signatures of nonequilibrium chemistry. The former was not detected
in the $L/M$-band spectra \citep{ske16,mor18,mil20}, but \citet{mil20}
reported the presence of weak CO absorption in the $M$-band spectrum.
In their modeling of the SED of WISE 0855, \citet{lac23} found that 
models with clouds and equilibrium chemistry provided the best fit.

Given that the SED of WISE 0855 has been sampled sparsely and typically with 
low SNRs and given the degeneracies and uncertainties in the model spectra, 
it has been unclear from modeling of previous data whether the atmosphere of 
WISE 0855 experiences H$_2$O ice clouds or nonequilibrium chemistry.

\subsection{Comparison of NIRSpec Data to Models}
\label{sec:models}

In an initial attempt at modeling the NIRSpec data for WISE 0855,
we have searched for model spectra that reproduce the PRISM data
from among available grids that are defined by the following features:
cloudless atmospheres with either chemical equilibrium and nonequilibrium 
chemistry \citep[{\tt ATMO 2020},][]{tre15,phi20,leg21,mei23};
cloudless atmospheres and chemical equilibrium \citep[Sonora Bobcat,][]{mar21}, 
cloudless atmospheres and nonequilibrium chemistry \citep{sau12}, 
partly cloudy atmospheres and chemical equilibrium \citep{mor12,mor14a},
and cloudy and cloudless atmospheres with chemical equilibrium and 
nonequilibrium chemistry \citep{lac23}. None of the grids provided
a sufficiently close match to the PRISM spectrum to be useful.

For all of the model suites that we considered, it is possible that the
parameters defining the grids are too sparsely sampled and that a good
fit could be achieved with custom models in which the parameters are tuned 
for WISE 0855. We have performed such an exercise with the {\tt ATMO 2020} 
models for both chemical equilibrium and nonequilibrium chemistry. 
As an initial model, we adopted the one for WISE 0855
from \citet{leg21}, which had $T_{\rm eff}=260$~K, a surface gravity
of log~g=4 [cm s$^{-2}$], an eddy diffusion coefficient of 
$K_{\rm zz}=10^{8.7}$, an effective adiabatic index of $\gamma=1.33$,
and solar values for metallicity and C/O.  We then increased the effective
temperature to improve the fit at 2--4~\micron, scaled all fluxes downward
to maintain agreement near 4.5--5~\micron\ (i.e., decreased the radius), and
decreased $\gamma$ to further improve the fit at $<$3~\micron.
The resulting model, which we refer to as the ``base model", has 
$T_{\rm eff}=285$~K and $\gamma=1.3$ while the remaining parameters are
unchanged from the model in \citet{leg21}.

\begin{figure}
\epsscale{1.1}
\plotone{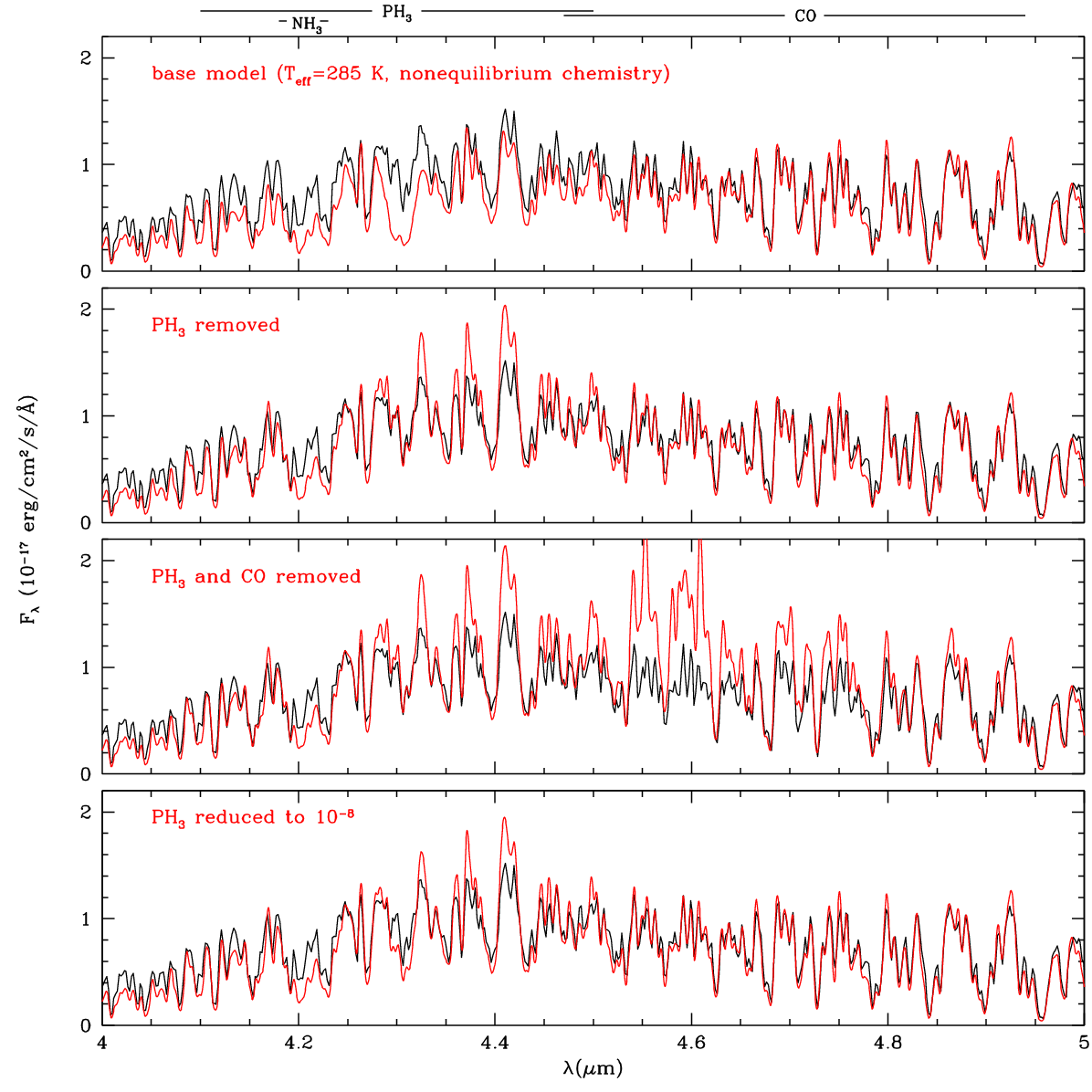}
\caption{NIRSpec G395M spectrum of WISE 0855 compared to spectra produced
by models of cloudless atmospheres \citep{tre15,phi20}. The top panel shows
our base model, which uses $T_{\rm eff}=285$~K and nonequilibrium chemistry. 
In the remaining panels, we have modified the base model to remove PH$_3$,
remove both PH$_3$ and CO, and reduce the mixing ratio of PH$_3$ to 10$^{-8}$.}
\label{fig:mod2} 
\end{figure}

In Figure~\ref{fig:mod2}, we show a portion of the G395M spectrum 
that includes bands from PH$_3$, NH$_3$, and CO.
The top panel compares the data to the spectrum from our base model. 
The prescription for nonequilibrium chemistry in that model produces mixing 
ratios of $7.7\times10^{-5}$ for NH$_3$ and $2\times10^{-7}$ for CO. 
The model does not account for mixing of PH$_3$, and instead uses the 
mixing ratio of $5\times10^{-7}$ derived for chemical equilibrium
\citep{phi20,leg21}. The strongest features from NH$_3$ and PH$_3$ are
evident in the model spectrum at 4.2 and 4.3~\micron, 
respectively. Additional absorption at other wavelengths from PH$_3$ and CO
is present in the model, but it does not appear as well-defined features 
against the backdrop of CH$_4$ and H$_2$O bands. 

To illustrate the influence of PH$_3$ and CO on the model spectra, we show in
Figure~\ref{fig:mod2} modified versions of the base model after removing 
the opacities of PH$_3$, removing both PH$_3$ and CO, and reducing the 
mixing ratio of PH$_3$ to $10^{-8}$. For each of these models, 
the pressure/temperature profile was reconverged to maintain 
hydrostatic and energy equilibria. The model with no PH$_3$
agrees well with the data near the wavelength of the strongest feature of
PH$_3$ (4.3~\micron), indicating that PH$_3$ is not detected.
The model that lacks both PH$_3$ and CO is significantly brighter than
the data across the wavelength range of CO (4.5--4.9~\micron), demonstrating
that CO is clearly detected. Meanwhile, the spectrum in that range is 
well-matched by the base model. The mixing ratio of CO in the base model 
is similar to the value derived by \citet{mil20} from modeling of the 
ground-based $M$-band spectrum obtained by \citet{ske16}.
In the model with a reduced mixing ratio of $10^{-8}$ for PH$_3$,
the 4.3~\micron\ feature is weak but detectable, which provides a 
rough upper limit on the abundance of that species.
In Figure~\ref{fig:mod2}, the models in which PH$_3$ has been 
removed or reduced are brighter than the data at 4.3--4.5~\micron, which is
discussed later in the context of the PRISM data.

As discussed in \citet{mil20}, the detection of CO in WISE 0855
indicates the presence of nonequilibrium chemistry due to vertical mixing.
The degree of vertical mixing needed to account for the observed
CO absorption in WISE 0855 is expected to produce strong PH$_3$ as well,
but it is not detected.
Like WISE 0855, the Y dwarf WISE 0359 exhibits absorption in CO but
not PH$_3$ in its NIRSpec data \citep{bei23}.

\begin{figure}
\epsscale{1}
\plotone{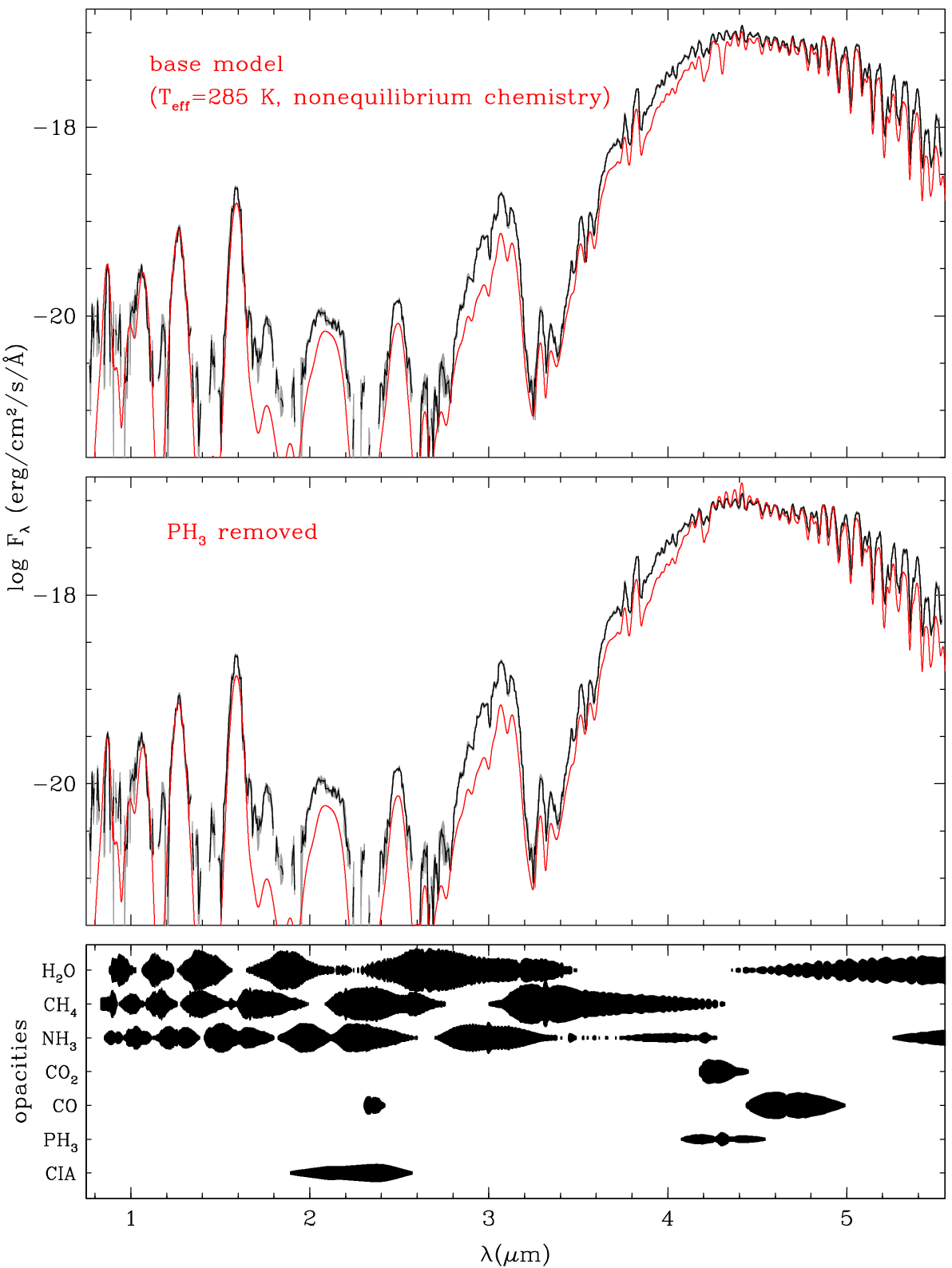}
\caption{NIRSpec PRISM spectrum of WISE 0855 compared to spectra produced
by models of cloudless atmospheres \citep{tre15,phi20}. The top panel shows
our base model for $T_{\rm eff}=285$~K and nonequilibrium chemistry.
In the bottom panel, we have modified the base model to remove PH$_3$.}
\label{fig:mod1}
\end{figure}

In Figure~\ref{fig:mod1}, we compare the PRISM data to the spectra
produced by the base model and the modified version with PH$_3$ removed,
which provided the best match to the G395M spectrum.
When normalizing the model to the data at 4.5--5~\micron\ as we have done,
the model agrees with the data at 0.8--1.6~\micron, is too faint 
at 1.6--4 and 5--5.6~\micron, and is too bright at 
4.3--4.5~\micron\ (see also Figure~\ref{fig:mod2}).
It is possible that the fit could be improved by adjusting the abundances
of some of the dominant absorbers. For instance, the abundance of
NH$_3$ in our model may need to be reduced based on the strength of the
4.2~\micron\ absorption feature (Figure~\ref{fig:mod2}).
The source of the discrepancy at 4.3--4.5~\micron\ is unclear.
We note that this wavelength range corresponds to the minimum in the
opacities of the dominant absorbers (Figure~\ref{fig:mod1}).

The temperature for our best model for WISE 0855 (285~K) is somewhat higher
than the value of 253--276~K derived by combining our luminosity estimate with
the luminosities predicted by evolutionary models (Section~\ref{sec:lum}).
As an alternative illustration of the difference, the model temperature
combined with our luminosity estimate correspond to a radius of
0.092~$R_\odot$, which is smaller than the radii of 0.10--0.11~$R_\odot$
that are predicted by evolutionary models for brown dwarfs at the luminosity
of WISE 0855 for ages of 1--10~Gyr.

As mentioned in Section~\ref{sec:comp}, \citet{ske16}, \citet{mor18},
and \citet{mil20} found that their ground-based $M$-band spectrum of
WISE 0855 was better matched by cloudy models than cloudless models, which 
was cited as evidence of H$_2$O ice clouds. However, the spectrum produced by
our cloudless model agrees well with the NIRSpec data in that wavelength
range (Figure~\ref{fig:mod2}).

Future studies can improve upon our modeling of WISE 0855 by 
incorporating 5--28~\micron\ data that were collected with
the Mid-infrared Instrument on JWST \citep[MIRI,][]{rie15} during the same
visit as the NIRSpec observations, optimizing the abundances of the dominant
absorbers in cloudless models like those from {\tt ATMO 2020}, tuning the 
latest cloudy models to fit WISE 0855 \citep{lac23}, and performing retrieval 
analysis \citep{bur17,lin17,zal19,row23}.

\subsection{Luminosity Estimate}
\label{sec:lum}

According to the model spectra for WISE 0855, the NIRSpec PRISM data
(0.77--5.53~\micron) should encompass roughly one third of its total flux. 
We have estimated the bolometric luminosity of WISE 0855 by 
integrating the flux in the PRISM spectrum and our 285~K model
at shorter and longer wavelengths and applying the parallactic distance
measured by \citet{kir21}, arriving at log~$L/L_\odot=-7.305\pm0.020$.
This calculation is similar to simply quoting the luminosity of our best fit 
model. As mentioned in Section~\ref{sec:comp}, the NIRSpec spectra have been
flux calibrated using the mean of multiple epochs of photometry 
at 4.5~\micron\ with Spitzer. We have adopted the standard deviation of
those measurements as the uncertainty in the flux calibration.
The error in our luminosity estimate does not include the uncertainty in
the model flux beyond the wavelength range of the PRISM data.
It will be possible to eventually eliminate most of that uncertainty by
incorporating the 5--28~\micron\ data from MIRI.

\begin{figure}
\epsscale{1.1}
\plotone{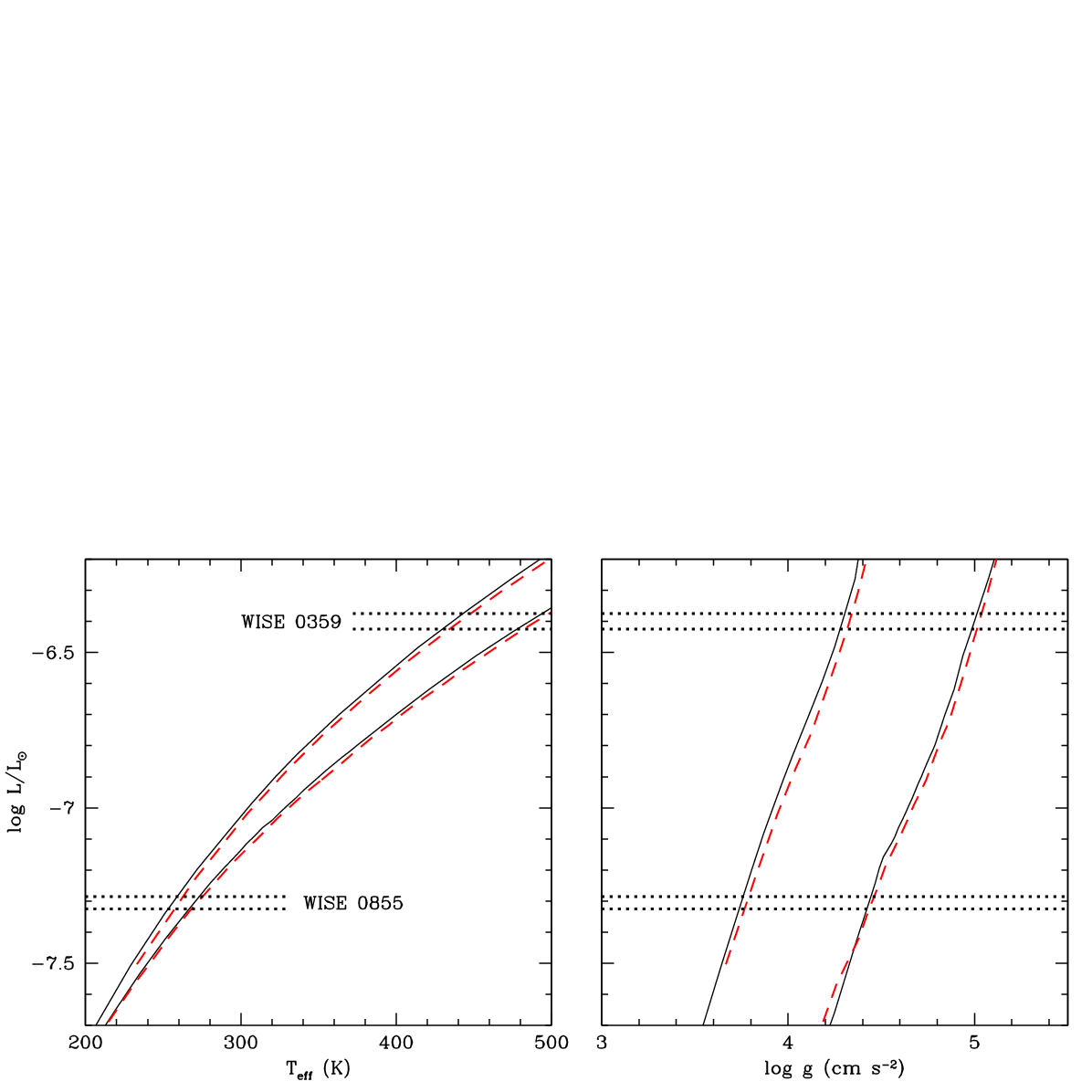}
\caption{Luminosity estimates for WISE 0855 (this work) and WISE 0359
\citep{bei23} compared to luminosities as a function of temperature and 
surface gravity predicted for ages for 1 and 10 Gyr by the evolutionary models 
of \citet[][solid lines]{mar21} and \citet[][dashed lines]{cha23}.}
\label{fig:lbol}
\end{figure}

We can estimate the effective temperature of WISE 0855 from its luminosity 
and the radii predicted by evolutionary models.
In Figure~\ref{fig:lbol}, we have plotted the luminosity estimate for
WISE 0855 with the luminosities as a function of temperature predicted
by two sets of evolutionary models for ages
of 1 and 10~Gyr, which encompass most members of the solar neighborhood.
We have included the luminosity for WISE 0359 from \citet{bei23} for comparison.
We have shown models of cloudless atmospheres with chemical equilibrium
and solar metallicity \citep{mar21,cha23}. Very similar luminosities
are produced by models with cloudy atmospheres and nonequilibrium chemistry 
\citep{sau12,mor14a,cha23}. Based on the evolutionary models, the luminosity 
of WISE 0855 corresponds to a temperature of 253--276~K for an age of 
1--10~Gyr, which is somewhat cooler than the value implied by the model spectra
(Section~\ref{sec:models}). As mentioned earlier, the true uncertainty
in our luminosity estimate is larger than the value we have quoted and plotted 
in Figure~\ref{fig:lbol}, so the same is true for the temperature estimate.

To illustrate the expected surface gravity of WISE 0855,
we have included in Figure~\ref{fig:lbol} a diagram of the predicted 
luminosities as a function of surface gravity for 1 and 10~Gyr.
The luminosity of WISE 0855 is indicative of log~$g\sim$4 [cm s$^{-2}$].
Meanwhile, the luminosity corresponds to a mass range of 3--10~$M_{\rm Jup}$ 
for ages of 1--10~Gyr according to the evolutionary models.

\section{Conclusions}

We have used NIRSpec on JWST to perform 1--5~\micron\ spectroscopy
on WISE 0855, which is the coldest known brown dwarf.
Our results are summarized as follows:

\begin{enumerate}

\item
We observed WISE 0855 in the fixed slit mode of NIRSpec with the
PRISM and G395M dispersers (R$\sim$40--300 and 1000), which produced useful
data at 0.77--5.53~\micron\ and 2.87--5.09/5.37--5.54~\micron, respectively.
NIRSpec has provided the first spectroscopy of WISE 0855 for most of those
wavelengths.
Although the brown dwarf is extremely faint at $<$2.5~\micron, the
combination of NIRSpec's sensitivity and the long exposure time for the
PRISM data (4.2~hrs) has yielded relatively high SNRs in that wavelength
range, with a median value of $\sim$20 for the brightest 50\% of pixels
(i.e., outside of the deepest bands).

\item
We have calculated synthetic photometry from the NIRSpec spectra in
several filters in which photometry has been previously measured.
The synthetic and imaging photometry are consistent with each other in
most filters with the exception of [3.6] from Spitzer.
The synthetic value of [3.6] from NIRSpec is $\sim$0.4~mag fainter than
the measurement from Spitzer, which implies that [3.6]$-$[4.5] is redder in
the spectrum than in the images. We find that the NIRSpec data for 
the Y0 dwarf WISE 0359 \citep{bei23} exhibits a similar discrepancy.
It would be useful to check whether the discrepancy extends to
other T and Y dwarfs observed by NIRSpec.

\item
We have performed preliminary modeling of the NIRSpec data
using the {\tt ATMO 2020} models of cloudless atmospheres \citep{tre15,phi20}.
Our best fitting model has $T_{\rm eff}=285$~K, 
nonequilibrium chemistry ($K_{\rm zz}=10^{8.7}$), an
effective adiabatic index of $\gamma=1.3$, and solar metallicity.
The modeling demonstrates a clear detection of the fundamental band
of CO, which is consistent with an earlier detection by \citet{mil20} with
ground-based data and indicates the presence of vertical mixing.
Previous observations of WISE 0855 and other Y dwarfs have found an
absence of PH$_3$ that seems inconsistent with the vertical mixing implied
by other species like CO \citep{mil20,bei23}. Similarly, NIRSpec does not
detect PH$_3$ in WISE 0855, providing an improved constraint in its
mixing ratio ($<10^{-8}$).

\item
The temperature of our best fitting model is $\sim$20~K higher than the
value derived when our luminosity estimate is combined with evolutionary models 
(i.e., the radius implied by the fit to the SED is somewhat smaller than
expected based on evolutionary models).

\item
Previous ground-based $M$-band spectroscopy of WISE 0855 (4.5--5.1~\micron)
has been cited for evidence of H$_2$O ice clouds, but we find that the NIRSpec
data in that wavelength range are matched well by our cloudless model.
Thus, clear evidence of H$_2$O ice clouds in WISE 0855 has not been 
identified yet, but it may still be present in the NIRSpec data.
The physical properties of WISE 0855, including the presence of H$_2$O clouds,
can be better constrained with more concerted modeling efforts using both
cloudless and cloudy models and the incorporation of 5--28~\micron\ data
from MIRI.

\end{enumerate}

\begin{acknowledgments}

We thank Michael Cushing and Samuel Beiler for helpful discussions and
Brianna Lacy for providing her model calculations.
P.T. acknowledges support from the European Research Council under grant
agreement ATMO 757858. R.P. acknowledges support from the Royal Society in
the form of a Dorothy Hodgkin Fellowship.  The JWST data were obtained from 
MAST at the Space Telescope Science Institute, which is operated by the 
Association of Universities for Research in Astronomy, Inc., under NASA 
contract NAS 5-03127. The JWST observations are associated with program 1230.
The Center for Exoplanets and Habitable Worlds is supported by the
Pennsylvania State University, the Eberly College of Science, and the
Pennsylvania Space Grant Consortium.

\end{acknowledgments}

\clearpage

\end{document}